\documentclass{article}
\usepackage{graphicx}

\usepackage{amsmath,amssymb}

\usepackage[dvips]{color}

\usepackage{latexsym,graphicx,epsfig}
\def\lb{\linebreak[4]}
\newcommand{\be}{\begin{equation}}
\newcommand{\ee}{\end{equation}}
\newcommand{\bea}{\begin{eqnarray}}
\newcommand{\eea}{\end{eqnarray}}
\newcommand{\bes}{\begin{subequations}}
\newcommand{\ees}{\end{subequations}}
\newcommand{\bear}{\begin{equation}\begin{array}}
\newcommand{\eear}[1]{\end{array}\label{#1}\end{equation}}
\newcommand{\fr}[2]{\dfrac{{ #1}}{{ #2}}}

\newcommand{\la}{\langle}
\newcommand{\ra}{\rangle}
\newcommand{\fn}[1]{\footnote{{#1}}}
\newcommand{\bu}{$\bullet$\ }
\renewcommand{\le}{\leqslant}

\def\vep{{\varepsilon}}

{\end{list}}
\newcounter{enumct}

\newcommand{\LL}[1]{\Lambda_{#1}}
\newcommand{\ls}[1]{\lambda_{#1}}
\newcommand{\ff}[2]{\left(\!\phi_#1^\dagger\phi_#2\!\right)}
\newcommand{\fft}[2]{\left(\!\Phi_#1^\dagger\Phi_#2\!\right)}

\newcommand{\fftc}[2]{\left(\!\!\Phi_#1^\dagger\Phi_#2\!\!-\!\!\fr{v^2}{2}\!\!\,\right)}

\newcommand{\Ls}[1]{\Lambda_{#1}}

\begin{document}
\renewcommand{\tilde}{\widetilde}

\date{}
\title{\Large\bf
A simple  criterium for  CP conservation in the most general 2HDM}
\author{I.~F.~Ginzburg,\\
\it Sobolev Institute of Mathematics, Novosibirsk, 630090, Russia;\\ \it Novosibirsk State University, Novosibirsk, 630090, Russia,\\
 M. Krawczyk,\\
\it Faculty of Physics, University of Warsaw, \\ \it ul. Pasteura 5, 02-093 Warsaw, Poland}

 \maketitle


\begin{abstract}
We find  set of  necessary and sufficient conditions for CP conservation in the most general 2HDM in terms of observable quantities. This set contains two relatively easily testable simple conditions instead of usually discussed three more complex  ones.

\end{abstract}

\section{Introduction}

CP violation is one of the important yet not well understood  problems in the fundamental physics.   Modern LHC data \cite{LHCdata} allow to conclude that the observed particle $h(125)$ is Higgs boson with spin-CP parity $0^{++}$ only under assumption that this particle has definite parity. Generally it can has no definite parity, as it happens in many models. In this case mentioned data give no information about  $h(125)$ parity \cite{Ginparity}.

In the Standard Model the CP violation is described by means of CKM matrix, but its origin remains unclear. The  extension of SM with two  Higgs doublets  instead of  one as in the SM,  called Two Higgs Doublet Model (2HDM),  has been   introduced in 1974  with the main aim for  providing  an extra source of CP violation \cite{Lee}. If this violation realizes, the physical spinless particles have no definite parity (and CP).

The 2HDM, as many models of New Physics beyond SM, contains many new particles and interactions.  A  problem arises how to check  in a simple way  whether this model  respects  CP symmetry. This is the problem which is  discussed in this paper.

\section{2HDM}
The 2HDM describes a system of two  spinless  isospinor fields $\phi_1$, $\phi_2$  with hypercharge $Y=1$.  The most general form of the 2HDM potential is
\bear{rl}
V\! & \!=\!\fr{\ls1}{2}\ff11^2+\fr{\ls2}{2}\ff22^2+\ls3\ff11\ff22         \\[3mm]
  & +\ls4\ff12\ff21 + \fr{\ls5}{2}\ff12^2 + \fr{\ls5^*}{2}\ff21^2   \\[3mm]
  & +\left[\ls6\ff11\ff12+\ls7\ff22\ff12+\text{h.c.}\right]         \\[3mm]
 \! & \! -\fr{m_{11}^2}{2}\!\ff11\!-\!\fr{m_{22}^2}{2}\!\ff22\!-\!\!\left[\fr{m_{12}^2}{2}\ff12\!+\!\!\text{h.c.}\right]\!
\eear{thdmpot}
Its coefficients are restricted by the requirement that the potential should   be positive at large quasiclassical values of $\phi_i$ ({\it positivity constraints}).  We assume also that these coefficients are not too big so that  one can use estimates based on  the lowest non-trivial approximation of the perturbation theory.

After EWSB the 2HDM  contains  3 neutral Higgs bosons $h_a\equiv h_{1,2,3}$, {in generally with indefinite CP parity,}  and charged Higgs boson $H^\pm$ with masses $M_a$ and  $M_\pm$, respectively\fn{The numbering of $h_a$ does not correspond  to an order of masses   $M_a$.}.

\subsection{Reparametrization freedom.} 2HDM describes system of two  fields with identical quantum numbers. Therefore, its description in terms of  original fields $\phi_i$ or  in terms of their linear superpositions $\phi'_i$   are equivalent;  this statement verbalizes the {\it reparameterization} (RPa) freedom of the model. The RPa group  consists of  RPa transformations
${\cal {\hat F}}$  of the form:
\bear{c}
\begin{pmatrix}\phi_1'\\ \phi_2'\end{pmatrix} =
\hat{\cal F}_{gen}(\theta, \tau,\rho)\begin{pmatrix}\phi_1\\ \phi_2\end{pmatrix}\,,\\[5mm]
\hat{\cal F}_{gen}=e^{i\rho_0}\begin{pmatrix}
\cos\theta\,e^{i\rho/2}&\sin\theta\,e^{i(\tau+\rho/2)}\\
-\sin\theta\,e^{-i(\tau+\rho/2)}&\cos\theta\,e^{-i\rho/2}
\end{pmatrix}.
\eear{reparam}
This transformation induces a transformation of the parameters of the Lagrangian
in such a way that the new Lagrangian, written in fields $\phi'_i$, describes the same physical content. We refer to these different  choices as - the different RPa bases.
A  subgroup of the RPa group -- the  rephasing  group RPh --  describes a  freedom  in  choice of the relative phase of fields $\phi_i$.

Transformation \eqref{reparam} is parameterized by  angles
$\theta,\, \rho,\,\tau$ and $\rho_0$. The parameter $\rho_0$  describes an overall phase transformation of the fields.
Since it does not affect the parameters of the potential,   it can be ignored.
The parameter $\rho$ describes the RPh symmetry   of system.

The {$U(1)_{EM}$ symmetry preserving} ground state of this system is given by a minimum of the potential
\be
\la\phi_1\ra =\begin{pmatrix}0\\ v_1/\sqrt{2}\end{pmatrix},\quad \la\phi_2\ra =\begin{pmatrix}0\\ v_2e^{i\xi}/\sqrt{2}\end{pmatrix},\,\label{ground}
\ee
with standard parameterization $v_1=v\cos\beta$, $v_2=v\sin\beta$.

\subsection{Higgs basis}   We use below   the RPa basis with $v_2=0$
(the Higgs, or Georgi, basis \cite{Georgi}), in which   the  2HDM potential  can be written in the form \cite{GKan}
\bear{c}
\!\!V_{HB}\! =\! M_\pm^2\fft22\! +\! \dfrac{\Ls1}{2}\fftc11^2\!\!
\!+\!\dfrac{\Ls2}{2}\fft22^2\\[2mm]
\!+\! \Ls3\fftc11\fft22\!+\!\Ls4\fft12\fft21\\[2mm]
       \! +\! \left[\dfrac{\Ls5}{2}\fft12^2\!+\!\Ls6\fftc11\fft12  
\! +\! \Ls7\fft22\fft12\!+\!\text{h.c.}\right].
\eear{HBmpot}
For this basis we use capital letters to denote  fields and parameters of potential, $\Phi_i$ and $\Ls  j$ respectively.

\subsection{Relative couplings }
In the discussion below we use the   relative couplings
for each neutral Higgs boson\fn{We  omit the adjective "relative" below.} $h_a\, (a=1,2,3)$:
\bear{c}
\chi^P_{a}=\fr{g^P_a}{g^P_{\rm SM} } \,,\quad
\chi^\pm_a=\fr{g(H^+H^-h_a)}{2M_\pm^2/v},\quad  \chi_{a}^{H^+ W^-} = \dfrac{g(H^+ W^- h_a)}{M_W/v}\,.
\eear{relcoupl}
 The quantities $\chi^P_{a}$ (where $P=V\,(W,Z),\;
q=t,b,...,\; \ell=\tau,...$) are the
ratios of the couplings of  $h_a$   with the fundamental particles $P$
to the corresponding  couplings for the would be SM Higgs boson with $M_h=M_a$.  The   other relative couplings describe interaction of $h_a$ with the charged Higgs boson   $H^\pm$. Couplings $\chi^V_a$ and $\chi^{\pm }_a$ are real due to Hermiticity of Lagrangian,   and are directly measurable. Couplings  $\chi_{a}^{H^+ W^-}$, $\chi^q_a$ and $\chi^\ell_a$ are generally complex.

There are useful sum rules among these couplings, namely
\be
(a)\;\sum\limits_a(\chi_a^V)^2=1\,,\quad
(b)\;(\chi_a^V)^2+|\chi_{a}^{H^+ W^-}|^2=1\,.\label{SR1}
\ee
Both real and imaginary parts of Yukawa couplings $\chi^q_a$ and $\chi^\ell_a$ can be measured in principle,   using  distributions of  Higgs bosons decay products\lb $h_a\to \bar{q}q$, $h_a\to \bar{\ell}\ell$. The absolute value of the  coupling $\chi_{a}^{H^+ W^-}$ is well measurable,  it  is fixed by the sum rule $(b)$.

   The unitarity of the rotation matrix  describing transition from
components of fields $\phi_i$ to the physical Higgs fields $h_a$, allows to obtain following relations  for couplings $\chi_a^{H^+ W^-}$ (the  factor $e^{i\rho}$   represents the   rephasing freedom in the Higgs basis):
\bear{cl}
\chi_1^{H^+ W^-}\equiv \left(\chi_1^{H^- W^+}\right)^*&=-   e^{i\rho}\;\fr{\chi_1^V\chi_2^V -i\chi_3^V}{\sqrt{1-(\chi_2^v)^2}},\\[2mm]
\chi_2^{H^+ W^-}\equiv \left(\chi_2^{H^- W^+}\right)^*&=e^{i\rho}\;\sqrt{1-(\chi_2^v)^2},\\[2mm]
\chi_3^{H^+ W^-}\equiv \left(\chi_3^{H^- W^+}\right)^*&=-e^{i\rho}\;\fr{\chi_2^V\chi_3^V +i\chi_1^V}{\sqrt{1-(\chi_2^v)^2}}.\eear{HWhcoupl}

\subsection{Minimal set of observables}

 In ref.~\cite{GKan} a minimal  complete
set of directly measurable quantities defining the 2HDM {\it (observables)}  was found
 ($a=1,2,3$):
\bear{l}
\mbox{\it v.e.v. of Higgs field  $v=246$~GeV};\\\mbox{\it masses of  Higgs bosons } M_a,\; M_\pm;\\
\mbox{\it 2 out of 3 couplings } \chi_a^V;\\
\mbox{\it 3 couplings } \chi^\pm_a;\\
  \mbox{\it quartic coupling } g(H^+H^-H^+H^-).
\eear{setpar}

In the most general 2HDM,  these observables are independent of each other. In  particular variants of 2HDM additional relations between these parameters may appear.

The parameters of potential in the Higgs basis are expressed through these  observables and  free parameter $\rho$, (appeared in $\LL {5,6,7}$ via   couplings  $\chi_a^{H^+ W^-}$):
\bear{llll}
\LL1 =&\sum\limits_a (\chi^V_{a})^2M_a^2/v^2\,;\quad &
 \LL4   =   &
    (\sum\limits_a M_a^2\!-\!M_{\pm}^2)/v^2\!-\!\LL1; \\[1mm]
 \LL5
 =&\sum\limits_a (\chi_{a}^{H^- W^+})^2M_a^2/v^2;\quad &
 \LL6 =&\sum\limits_a \chi^V_{a}\chi_{a}^{H^- W^+}M_a^2/v^2;\\[2mm]
\LL3=& 2\left(M_\pm^2/v^2\right)\sum\limits_a \chi_{a}^V\chi_a^\pm;\quad &
 \LL7=&
2\left(M_\pm^2/v^2\right)\sum\limits_a \chi_{a}^{H^-W^+}\chi_a^\pm\,;\\[2mm]
\LL2= &  2g(H^+H^-H^+H^-)\,.&&
 \eear{Mextract}

The parameters of the potential \eqref{thdmpot} with  the defined  values of $\tan\beta$ and $\xi$ \eqref{ground} are obtained from parameters \eqref{Mextract} with the aid of transformation \eqref{reparam} of the form
\bear{c}
    \begin{pmatrix}{\phi}_1\\ \phi_2\end{pmatrix}\! =\!
    \hat{\cal F}_{HB}\!\begin{pmatrix}\Phi_1\\ \Phi_2\end{pmatrix},\;\; \hat{\cal F}_{HB}\!=\! \hat{\cal F}_{gen}(\theta\!=\!-\beta, \tau\!=\!\xi, -\rho).
\eear{reparamHB}

\section{{Conditions}  for a CP conservation}

In fact,  the C-parity is  not defined for the system of spinless particles, for such  system only P-parity is defined. When in addition we consider
 fermions, the P-parity violation is transformed to the CP violation
(see e.g. \cite{hunter1,hunter2}).

In the CP violated case a transition like $h_a\to\bar{q}_1 q_1$leads to the  fermion state with indefinite CP parity (mixture of  CP-even and CP-odd components). Therefore, transitions $\bar{q}_1 q_1\to h_a\to \bar{q}_2 q_2$ violate the CP  symmetry and  CP-odd states can be transformed into CP-even ones and vice versa.  This very opportunity is treated as the CP violation.

\subsection{General  observations}

The CP symmetry  is conserved in some  model  containing Higgs bosons if
\bes\label{critCP}
\bea
\begin{array}{l}
\mbox{\bu \it Each observable physical neutral  spinless  Higgs boson  has }\\
\mbox{\it definite P-parity, in 2HDM -- P-even $h_1, h_2$ and P-odd $h_3$.}\end{array}\label{critCPI}\\
\mbox{\bu \it There are no P-violating interactions between these scalars.}
\label{critCPII}
\eea\ees
In the standard notations used for 2HDM,  P-even scalars  $h_1, \,h_2$ are denoted
$h,\,H$ and  P-odd  $h_3$ is called $A$.  Point \eqref{critCPII} means  an absence of  the  interactions
\be
h_ih_jh_3,\quad h_3h_3h_3,\quad h_ih_jh_kh_3,\quad h_ih_3h_3h_3 \quad (i,j,k=1,2).\label{CPoddint}
\ee

It is well known that in the 2HDM this criterium is fulfilled if
\bes\label{critCPBas}
\bea
\mbox{\it The RPa basis exists in which:\;\;\;\.\;\;\; }\nonumber\\
\mbox{\bu \it all parameters  of potential
are real};\label{critCPBasa}\\
\mbox{\bu \it relative phase \eqref{ground} $\xi= 0$.\,\;\;\;\.\;\;\; \;\;\;\.\;\;\; }\label{critCPBasb}
\eea\ees

In is worth mentioning,  that the condition \eqref{critCPBasa} forbids an explicit CP violation while  conditions \eqref{critCPBasa}  and \eqref{critCPBasb} {together} forbid a  spontaneous CP violation.

The equation  \eqref{critCP}  only  describes CP-conservation, but does not  provide  a criterium for CP conservation (or violation).   The description
\eqref{critCPBas} is RPa  basis-dependent.

\subsection{Method of the CP-odd  basis-independent invariants}

To obtain  criterium  for CP violation many authors  constructed the RPa basis-invariant  CP-odd combinations of parameters of the Higgs potential.
Then a condition for  CP conservation is formulated as a demand of vanishing of  all these invariants.
For 2HDM three such  invariants $Im{\cal J}_{1,2,3}$ were found   in ref.~\cite{QQQ1}, for multi-Higgs models such invariants were constructed in ref.~\cite{multi}.

The invariants \cite{QQQ1} for  2HDM were expressed via  measurable quantities  in ref.~\cite{QQQ}.    In the terms of quantities  \eqref{setpar}, the  corresponding conditions for the CP conservation read as
\bear{ll}
Im {\cal J}_1&=\sum\limits_{i,j,k}\vep_{ijk}\fr{2M_i^2M_\pm^2}{v^4}\chi_i^V \chi_k^V\chi_j^\pm=0\,,\\[3mm]
Im{\cal J}_2 & =2\chi_1^V\chi_2^V\chi_3^V\sum\limits_{i,j,k}\vep_{ijk}\fr{M_i^4M_k^2}{v^6}=0\,,\\[3mm]
Im{\cal J}_{30}&=4\sum\limits_{i,j,k}\vep_{ijk}
\fr{(M_\pm^2\chi_i^\pm+ M_i^2\chi_i^V)M_i^2M_\pm^2}{v^6}\chi_j^V\chi_k^\pm=0\,.
\eear{GOOparam}

Note that in the discussed approach there are four CP-odd invariants and one should check vanishing only of  two of them (see e.g. \cite{hunter2}). Since the choice of these two is not fixed from beginning, the presented set contains three conditions, instead of necessary two.
Besides, in our opinion   the equations \eqref{GOOparam} are   too complicated and    their experimental  verification, discussed in \cite{QQQ},  requires   too complex procedure.

\subsection{A  direct criterium { for CP conservation}}\label{secdirect}

We   formulate  conditions for CP conservation without  using an intermediate
CP-odd RPa-invariants.  We start with a   description of CP conservation \eqref{critCP} and use only observables.\\

In Refs.~\cite{Gin14}, \cite{Gin15-1} we describe limitation for CP violation, based only on one condition\fn{Particular version of such approach was used in \cite{Santos15}.}
\eqref{critCPI}, without checking up of condition  \eqref{critCPII},  in the form
\be
\;\prod\limits_a\chi_a^V=0\,,\quad \; \prod\limits_a\chi_a^\pm=0\,,\quad \;\left|\prod\limits_a \chi^f_a\right| =\prod\limits_a \left|\chi^f_a\right|\,.\label{CPinit}
\ee
 Below we
simplify these conditions and to prove that the set of new conditions is {\it necessary and sufficient} one.\\

\noindent { \bf \bu The  direct criterium}.\\

In the CP conserved case all $h_a$ should have definite parity. In particular, one of them is P-odd,  while two others are P-even \eqref{critCPI}.
Therefore, the necessary condition for a CP conservation is an existence of one neutral  Higgs boson (we denote it  $h_3$),  which doesn't couple to the CP-even states $VV$ and $H^+H^-$:
\be
 \boxed{\begin{array}{c}
 \mbox{There exists a  neutral Higgs boson $h_3$ for which}\\
 g(h_3VV)=0\,,\qquad g(h_3H^+H^-)=0\,.\end{array}}\label{crit1}
\ee

Now, one has to check condition \eqref{critCPII}.  In order to do this, we substitute Eq-s. \eqref{crit1}, \eqref{HWhcoupl} into  \eqref{Mextract},   choosing $\rho=0$. One can see that all  parameters of potential $\Ls a$ in the Higgs basis    are real.  Therefore, in view of a statement \eqref{critCPBas}, the CP-symmetry  of model is not violated.  In particular, the  CP violated interactions \eqref{CPoddint} don't appear.  (Besides, it is easy to check that conditions
\eqref{crit1} ensure compliance of conditions  \eqref{GOOparam} and first two conditions \eqref{CPinit}.)
  Therefore we conclude  that {\bf the conditions  \eqref{crit1}
are  necessary and sufficient}  for establishing  CP conservation in the 2HDM.\\

\noindent {\bu\bf  A discussion of a direct criterium. }\\
It is worth to note, that the condition $g(h_3VV)=0$ means that  the scalar $h_3$  has no P-even admixture. This results  immediately  to the   identities $g(h_3h_kZ)=0$ ($k=1,2$).  This conclusion allows to replace checking up of the condition $g(h_3VV)=0$  by checking of one of conditions $g(h_3h_kZ)=0$, i.~e. non-observation of corresponding decays, as it was proposed\fn{Let us remind that  such condition (including all its forms discussed in \cite{Santos15})  is insufficient to establish  CP conservation in the considered  model in the general case, since does not guarantee the fulfillment of the condition \eqref{critCPII}, i.~e. non-appearance of CP odd vertices \eqref{CPoddint}}
 in  \cite{Santos15}.
\begin{figure}[htb]
\begin{center}
\includegraphics[width=0.3\textwidth,height=2.5cm]{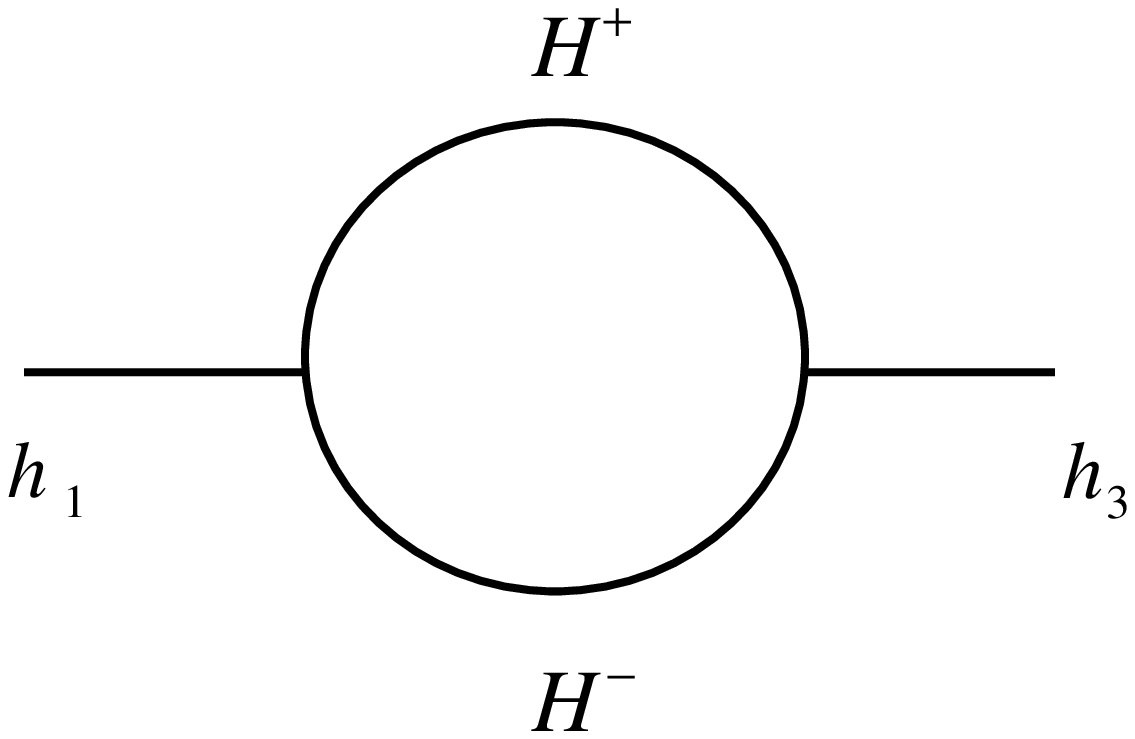}\hspace{20mm}\vspace{2mm}
\includegraphics[width=0.3\textwidth,height=2.5cm]{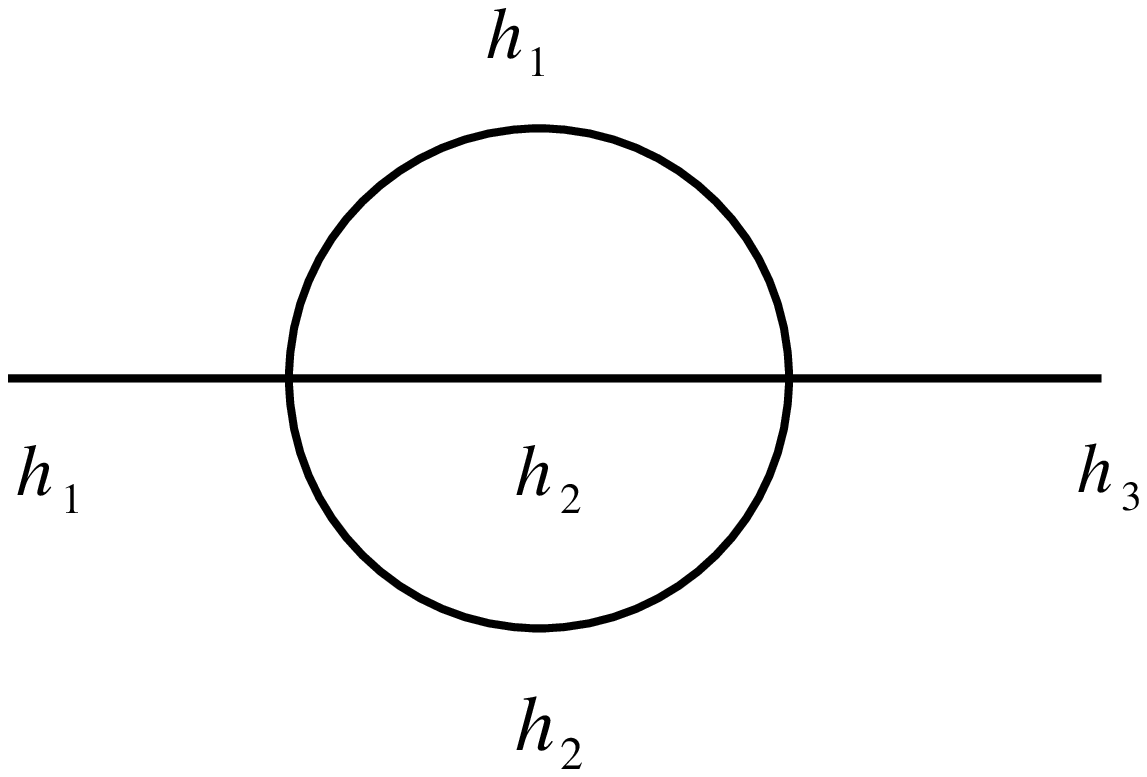}\vspace{-4mm}
\caption{\it Mixing obliged by the $H^+H^-$ loop (left) and one of CP violated vertices \eqref{CPoddint} (right).} \end{center}\label{fig:mixing}
\end{figure}\\[-6mm]
If $g(h_3H ^+H^-)\neq 0$, the loop diagrams  of  Fig.1 mix the  Higgs states with different incident CP parity, resulting in CP breaking.\\

\noindent {\bf \bu  Consequence for the Yukawa interaction}.

The Yukawa interaction
can violate CP symmetry of the model only in the case of a P non-conservation for scalars.
The  conditions \eqref{crit1} guarantee that $h_a$ are pure state of CP parity ($h_1$ and $h_2$ are CP even, $h_3$ is CP odd). In this case Yukawa interactions cannot generate CP violation, and the third  condition \eqref{CPinit} is fulfilled automatically.

Certainly,  interactions of fermions, different from the Yukawa  interaction with Higgs particles, can violate CP due to some other mechanisms; for example the CKM matrix describes such \underline{outer} violation. Such violation can be transferred to the Higgs sector,  presumably  as a small correction. To  observe  this kind {\it induced} CP violation, the last condition of \eqref{CPinit} should be checked
\be
\left|\prod\limits_a \chi^f_a\right| =\prod\limits_a \left|\chi^f_a\right| \;\; \mbox{\it for each fermion $f$}\,. \label{CPf}
\ee

\section{Possibilities for  a verification}

The verification of CP conservation requires an  observation of all scalars  of the model. In the realized   in Nature SM-like scenario  this  looks  difficult (see e.g. \cite{Gin15-1}). Moreover, one should check  that some measurable quantities are equal 0. In any case,  these measurements cannot pretend for a high accuracy. From this point of view the proposal to change a direct criterium  to a condition for  a non-observation of decay $h_3\to h_1Z$, etc. given  in \cite{Santos15}  looks  attractive. Nevertheless,  one  cannot expect high accuracy in the statement  about CP conservation in the 2HDM.

One can imagine the situation in which CP violation is concentrated in the coupling $g(h_3H^+H^-)\neq 0$ and interactions \eqref{CPoddint} while $g(h_3VV)$ and  $g(h_3h_kZ)$ are small  (e.~g. if they appear only at the loop level).  In such case,  due to experimental inaccuracy, measuring  only  the latter couplings  can not  give information about CP violation in the model.

\section{Conclusions}

We present here the new,   necessary and sufficient conditions for CP conservation in 2HDM \eqref{crit1},  which are common for all mechanisms of CP violation. We prove that the verification of CP conservation in 2HDM requires to measure two simple and  relatively easily testable  observables  instead of three more complex   conditions  \eqref{GOOparam} discussed by many authors \cite{QQQ}, \cite{QQQ1}.

\section*{Appendix. Necessary condition for { CP conservation in} Multi Higgs Doublet Model}

The  criterium of CP conservation in the Multi Higgs Doublet Models -- nHDM are also of interest. Here, the method of CP-odd invariants { allows to construct many equations, which can be used for obtaining  conditions for CP conservation. Both complete set of these equations and their expressions via measurable quantities are absent up to now (see e.~g. \cite{multi}).}

The direct method  { used above} allows to formulate  for the nHDM   simple necessary conditions  for the CP conservation.

After EWSB the  nHDM  contains  $2n-1$ neutral Higgs bosons $h_a$, generally with indefinite CP parity,  and $n-1$ charged Higgs boson $H_b^\pm$ with masses $M_a$, $M_{b\pm}$, respectively. The couplings $h_aVV$ obey { first} sum rule \eqref{SR1}. In the case of CP conservation one can  split
 spinless neutral particles $h_a$ into two groups:  P-even $h_1,...,h_n$ and P-odd $h_{n+1},...,h_{2n-1}$. Similarly  to \eqref{crit1}, the condition for a CP conservation in the nHDM can be written as
\be
\boxed{\begin{array}{l}\mbox{ There exist  $n-1$ neutral  Higgs bosons $h_c$, for which}\\
g(h_cVV)=0\,,\qquad g(h_cH_b^+H_b^-)=0\,\quad (n+1\le c\le 2n-1).\end{array}}\label{critmulti}
\ee
These $n(n-1)$ conditions are {\it necessary} for CP conservation. We don't know now whether these conditions are {\it sufficient} or not.\\

\section*{Acknowledgments}
We are thankful to G.~C. Branco, I.P. Ivanov, M. Rebelo,  R. Santos   for discussions. This work was supported in part by grants RFBR 15-02-05868,
NCN OPUS 2012/05/B/ST2/03306 (2012-2016) and HARMONIA project under contract
  UMO-2015/18/M/ST2/00518 (2016-2019).\\


\end{document}